\documentclass[manuscript]{aastex}
\usepackage{txfonts}
\usepackage{latexsym,bm}
\usepackage{lineno}
\usepackage{multirow}
\shorttitle{Simulation of gradual SEP events}

\shortauthors{Wang et al.}

\begin{document}

\title{A numerical simulation of solar energetic particle dropouts during impulsive events}

\author{ Y. Wang\altaffilmark{1}, G. Qin\altaffilmark{1},
M. Zhang\altaffilmark{2} and S. Dalla\altaffilmark{3}}

\email{ Yang Wang,ywang@spaceweather.ac.cn; Gang Qin, gqin@spaceweather.ac.cn}


\altaffiltext{1}{State Key Laboratory of Space Weather, National Space Science Center, Chinese Academy of Sciences,
Beijing 100190, China}
\altaffiltext{2}{Department of Physics and Space Science,
Florida Institute of Technology, Melbourne, Florida 32901, USA}
\altaffiltext{3}{Jeremiah Horrocks Institute, University of Central 
Lancashire, Preston, Lancashire PR1 2HE, UK}


\begin{abstract}
This paper investigates the conditions for producing rapid variations of solar energetic
particle (SEP) intensity commonly known as ``dropouts''. In particular, we use numerical
model simulations based on solving the focused transport equation in the 3-dimensional Parker
interplanetary magnetic field to put constraints on the properties of particle transport coefficients
both in the direction perpendicular and parallel to the magnetic field.
Our calculations of the temporal intensity profile of 0.5 and 5 MeV protons at the Earth show that
the perpendicular diffusion must be small enough while the parallel mean free path should
be long in order to reproduce the phenomenon of SEP dropouts. When the parallel
mean free path is a fraction of 1 AU and the observer is located at $1$ AU, the perpendicular to parallel diffusion ratio
must be below $10^{-5}$, if we want to see the particle flux 
dropping by  at least several times within three hours. 
When the observer is located at a larger solar radial distance, the perpendicular to parallel diffusion ratio for reproducing the dropouts should be even lower than that in the case of 1 AU distance.
A shorter parallel mean free path or a larger radial distance from the source to observer will cause the particles to arrive later,
making the effect of perpendicular diffusion more prominent and SEP dropouts disappear. All these effects require that the magnetic turbulence that resonates with the particles must be low everywhere in the inner heliosphere.



\end{abstract}


\section{INTRODUCTION}

Solar Energetic Particles (SEPs) encounter small-scale irregularities during transport in the large scale interplanetary magnetic field. The particles are scattered by the
irregularities whose scales are comparable to the particles' gyro radius. 
The parallel diffusion is produced by the pitch angle scattering, while
the perpendicular diffusion is caused by crossing the local field line or following  magnetic field lines randomly walking in space. Low-rigidity particles tend to 
follow more tightly  along individual field lines, whereas high-rigidity particles can 
cross local field lines  
more easily. As a result, the perpendicular diffusion of lower rigidity particles is
generally smaller than that of high-rigidity particles.

The  diffusion coefficients  of SEPs  depend on the magnetic turbulence in
the solar wind.  Jokipii (1966) was the first to use the Quasi-Linear Theory (QLT) to calculate particle diffusion coefficients from magnetic turbulence spectrum. But later it was found that the observed particle
mean free paths are usually much larger than the QLT results derived from a slab magnetic turbulence \citep{Palmer82}. 
According to 
\cite{matthaeus1990evidence}, the slab model is not a good approximation to describe
 the Interplanetary Magnetic Field (IMF) turbulence because there is also a stronger  two-dimensional (2D) component. \citet{BieberEA94} showed that with a ratio of 
turbulence energy between slab and 2D components, $E^{slab}:E^{2D}=20:80$,   the QLT was able to derive a parallel mean free path much better in agreement with observations.
However, the perpendicular diffusion remained a puzzle for many years. It is shown 
that the particles' perpendicular diffusion model of Field Line Random Walk (FLRW) 
based on QLT has difficulty to describe spacecraft observations and numerical 
simulations. Recently, the Non-Linear Guiding Center Theory (NLGC) 
\citep{Matthaeus2003ApJ...590L..53M} was developed to describe the perpendicular 
diffusion in magnetic turbulence.
The perpendicular diffusion coefficient from the NLGC theory  agrees quite 
satisfactorily with numerical simulations of particle transport in typical solar 
wind conditions.

Observations by the ACE and Wind spacecraft show that there are rapid temporal structures 
in the time profiles of $ \sim 20 $ keV nucleon$^{-1}$ to $ \sim 5 $ 
MeV nucleon$^{-1}$  ions during impulsive SEP events.
The  phenomenon is commonly known as ``dropouts''  or
``cutoffs'' in some cases.
In the dropouts, the particle intensities exhibit short time scale (about several hours) variations,
whereas, the cutoffs are referred to as some special dropouts
in which the intensities suddenly decrease without recovery.
They do not seem to be associated with visible local magnetic field changes
\citep{mazur2000interplanetary,Gosling2004Correlated,chollet2008multispacecraft,Droge2010}.
Contrarily to the previous
studies, by performing a detailed analysis of magnetic field topology
during SEP events, \cite{trenchi2013solar,trenchi2013observations} identified magnetic structures associated with SEP
dropouts. \cite{,trenchi2013solar} found that SEP dropouts are generally
associated with magnetic boundaries which represent the borders between
adjacent magnetic flux tubes while \cite{trenchi2013observations}, using the Grad-Shafranov reconstruction, identified flux ropes or current sheet
associated with SEP maxima.
The dropouts and cutoffs can be interpreted  as a result of magnetic field lines that connect or disconnect the observer alternatively to the SEP source on the Sun.
However, with perpendicular diffusion, particles can cross the field 
lines as they propagate in the interplanetary  
space. A strong enough perpendicular diffusion can efficiently diminish longitudinal gradients of fluxes. The dropouts and cutoffs 
provide a good chance for us to estimate the level of perpendicular diffusion in the interplanetary space.

In an effort to interpret the SEP dropouts, \cite{giacalone2000small} did a simulation of
test particle trajectory in a model with random-walking magnetic field lines
using Newton-Lorentz equation to study SEPs
dropouts. It was found that the phenomenon is consistent with random-walking
magnetic field lines. In addition, it was found from their simulations that,  particle perpendicular diffusion relative to the Parker spiral due to the field line random walk can be significant, and
the ratio of perpendicular diffusion to 
the parallel one relative to the Parker spiral can be as large as $2\%$. However, their perpendicular diffusion coefficients relative to the background magnetic field (instead of Parker spiral) could  still be very small, so dropouts can be obtained.
Recently,  using the same technique as in \cite{giacalone2000small},  \cite{Guo2013Small}   found that in some condition, dropouts  can be 
reproduced in the foot-point random motion model, but no dropout is  seen in the slab + 2D model. 
In \cite{Droge2010}, the large scale magnetic field is assumed to be a Parker 
spiral, and the observer is located at $1$ AU equatorial plane.
At the start the observer is connected to the source region, and leaves the region 
after some time due to the effect of co-rotation.
Based on a numerical solution of the focused transport equation,  
they found that in order 
to reproduce cutoffs, the ${\kappa _ \bot }/{\kappa _\parallel }$ should be as small as a few times of ${10^{ - 5}}$.  In the turbulence view, some other mechanisms \citep{ Ruffolo2003Trapping, Chuychai2005Suppressed,  Chuychai2007Trapping, Kaghashvili2006Propagation,  Seripienlert2010Dropouts} are also  proposed to interpret the dropout phenomenon.

In this work, we use a Fokker-Planck focused transport equation to calculate the transport of SEPs in three-dimensional Parker interplanetary magnetic field. We intend to  put constraints on the conditions of the perpendicular and parallel diffusion coefficients 
for observing the SEP dropouts and cutoffs.
In SECTION 2 we describe our SEP transport model. In SECTION 3
simulation results are presented. In SECTION 4 the simulation results are 
discussed, and conclusions based on our simulations are made.

\section{MODEL}
Our model is based on solving a  three-dimensional focused transport equation following the same method in our previous research \citep[e.g.,][]{Qin2006JGRA..11108101Q,Zhang2009ApJ...692..109Z,Wang2012apj}.
The transport equation of SEPs can be written as
\citep{Skilling1971ApJ...170..265S,Schlickeiser2002cra..book.....S,
Qin2006JGRA..11108101Q,Zhang2009ApJ...692..109Z,Droge2010,HeEA11,Wang2012apj,
zuo2013acceleration,qin2013transport}
\begin{eqnarray}
  \frac{{\partial f}}{{\partial t}} = \nabla\cdot\left( \bm
  {\kappa_\bot}
\cdot\nabla f\right)- \left(v\mu \bm{\mathop b\limits^ \wedge}
+ \bm{V}^{sw}\right)
\cdot \nabla f + \frac{\partial }{{\partial \mu }}\left(D_
{\mu \mu }
\frac{{\partial f}}{{\partial \mu }}\right) \nonumber \\
  + p\left[ {\frac{{1 - \mu ^2 }}{2}\left( {\nabla  \cdot \bm{V}^
  {sw}  -
\bm{\mathop b\limits^ \wedge  \mathop b\limits^ \wedge } :\nabla
\bm{V}^{sw} } \right) +
\mu ^2 \bm{\mathop b\limits^ \wedge  \mathop b\limits^ \wedge}  :
\nabla \bm{V}^{sw} }
\right]\frac{{\partial f}}{{\partial p}} \nonumber \\
  - \frac{{1 - \mu ^2 }}{2}\left[ { - \frac{v}{L} + \mu \left
  ( {\nabla  \cdot
\bm{V}^{sw}  - 3\bm{\mathop b\limits^ \wedge  \mathop b\limits^
\wedge}  :\nabla \bm{V}^{sw} }
\right)} \right]\frac{{\partial f}}{{\partial \mu }},\label{dfdt}
\end{eqnarray}
where $f(\bm{x},\mu,p,t)$ is the gyrophase-averaged  particle distribution function  as a function of time $t$, position in a non-rotating heliographic coordinate system
$\bm{x}$, particle momentum $p$ and pitch angle cosine $\mu$ in the
plasma reference frame. In the equation, $v$ is the particle speed,
$\bm{\mathop b\limits^ \wedge}$ is a unit vector
along the
local magnetic field; 
$\bm{V}^{sw}=V^{sw}\bm{\mathop r\limits^ \wedge}$ is the solar
wind velocity in the radial direction; and $L$ is the magnetic focusing length given
by
$L=\left(\bm{\mathop b\limits^ \wedge}\cdot\nabla\\{ln} B_0
\right)^{-1}$ with $B_0$ being the magnitude of the background  IMF. This equation includes many important particle transport effects such as particle streaming along field line, adiabatic cooling, magnetic focusing,  and the diffusion coefficients parallel and perpendicular to the IMF.
It is noted that for low-energy SEPs propagating  in inner heliosphere, the drift effects can be neglected.
Here, we use the Parker field model for the IMF, and the solar wind speed is  $400$ km/s.

The parallel particle
mean free path  $\lambda _\parallel$ is related to the particle pitch angle
diffusion $D_{\mu\mu}$ through \citep{Jokipii1966ApJ...146..480J,Earl74}
\begin{equation}
\lambda _\parallel   = \frac{{3v}}{8}\int_{ - 1}^{ + 1}
{\frac{{\left(1 - \mu ^2 \right)^2 }}{{D_{\mu \mu } }}d\mu
}\label{lambda_parallel_1},
\end{equation}
and parallel diffusion coefficient $\kappa_\parallel$ can be written as 
$\kappa_\parallel=v\lambda_\parallel/3$.

We choose to use a pitch angle diffusion coefficient from \citep{Beeck1986ApJ...311..437B, QinEA05, 
Qin2006JGRA..11108101Q}
\begin{equation}
{D_{\mu \mu }}(\mu ) = {D_0}v{R^{s - 2}}\left({\mu ^{s - 1}} + h\right)\left(1 - 
{\mu ^2}\right)
\end{equation}
where the constant ${D_0}$ is adopted from \cite{Teufel2003A&A397}
\begin{equation}
{D_0} = {\left( {\frac{{\delta {B_{slab}}}}{{{B_0}}}} \right)^2}\frac{{\pi (s - 1)}}{{4s}}{k_{\min }}
\end{equation}
here  $ \delta {B_{slab}} $ is  the magnitude of slab turbulence, 
$ {k_{\min }} $ is the lower limit of wave number of the inertial range in the slab turbulence power spectrum, 
$ R = pc/\left( {\left| q \right|{B_0}} \right) $ is the maximum particle Larmor radius, $q$ is the  particle charge, and
$s = 5/3$ is the Kolmogorov spectral index of the magnetic field turbulence in the inertial range. The constant $h$ comes from the non-linear effect of magnetic turbulence on the pitch angle diffusion at $\mu=0$ \citep{QinAShalchi09,QinAShalchi14}.
In following simulations, we set  $ h=0.01$,  and ${k_{\min }} =1/l_{slab}$, where, $l_{slab}$ is the slab turbulence correlation length. 
In this formula, we assume that ${\left( {\delta {B_{slab}}} \right)^2}/{\left( {{B_0}} 
\right)^2} \cdot {k_{\min }} = {A_1}$. Different parallel particle mean free path 
values can be obtained by altering the parameter $A_1$.

The perpendicular diffusion coefficient is taken from the NLGC theory
\citep{Matthaeus2003ApJ...590L..53M} with the following
analytical approximation \citep{ShalchiEA04,ShalchiLi2010}
\begin{equation}
{\bm{\kappa} _ \bot } = \frac{1}{3}v{\left[ {{{\left( {\frac{{\delta {B_{2D}}}}{{{B_0}}}} \right)}^2}\sqrt {3\pi } \frac{{s - 1}}{{2s}}\frac{{\Gamma \left( {\frac{s}{2} + 1} \right)}}{{\Gamma \left( {\frac{s}{2} + \frac{1}{2}} \right)}}{l_{2D}}} \right]^{2/3}}{\lambda _\parallel }^{1/3}\left( {{\bf{I}} - \bm{ \mathop b\limits^ \wedge} \bm{ \mathop b\limits^ \wedge}  } \right)  
\end{equation}
where $B_{2D}$ and $ l_{2D} $ are the magnitude and the correlation
length of 2D component of magnetic turbulence, respectively. $ \Gamma $ is the gamma function.
Here for simplicity, $\bm{\kappa _\bot}$ is assumed to be
independent of $\mu$, since particle pitch angle diffusion  usually is much faster than  the perpendicular diffusion, so 
the particle sense the effect of perpendicular diffusion averaged over all pitch angles. $\bm{{\rm I}}$ is
a unit tensor. In our simulations, we set
${\left( {\delta {B_{2D}}} \right)^2}/{\left( { {B_{0}}} \right)^2} \cdot{l_{2D}}= A_2$, and $s = 5/3$. As a result,  the value of
perpendicular diffusion coefficient can be altered by
changing the parameter  $A_2$, and parallel diffusion coefficient. 

\begin{figure}
\epsscale{.50}
\plotone{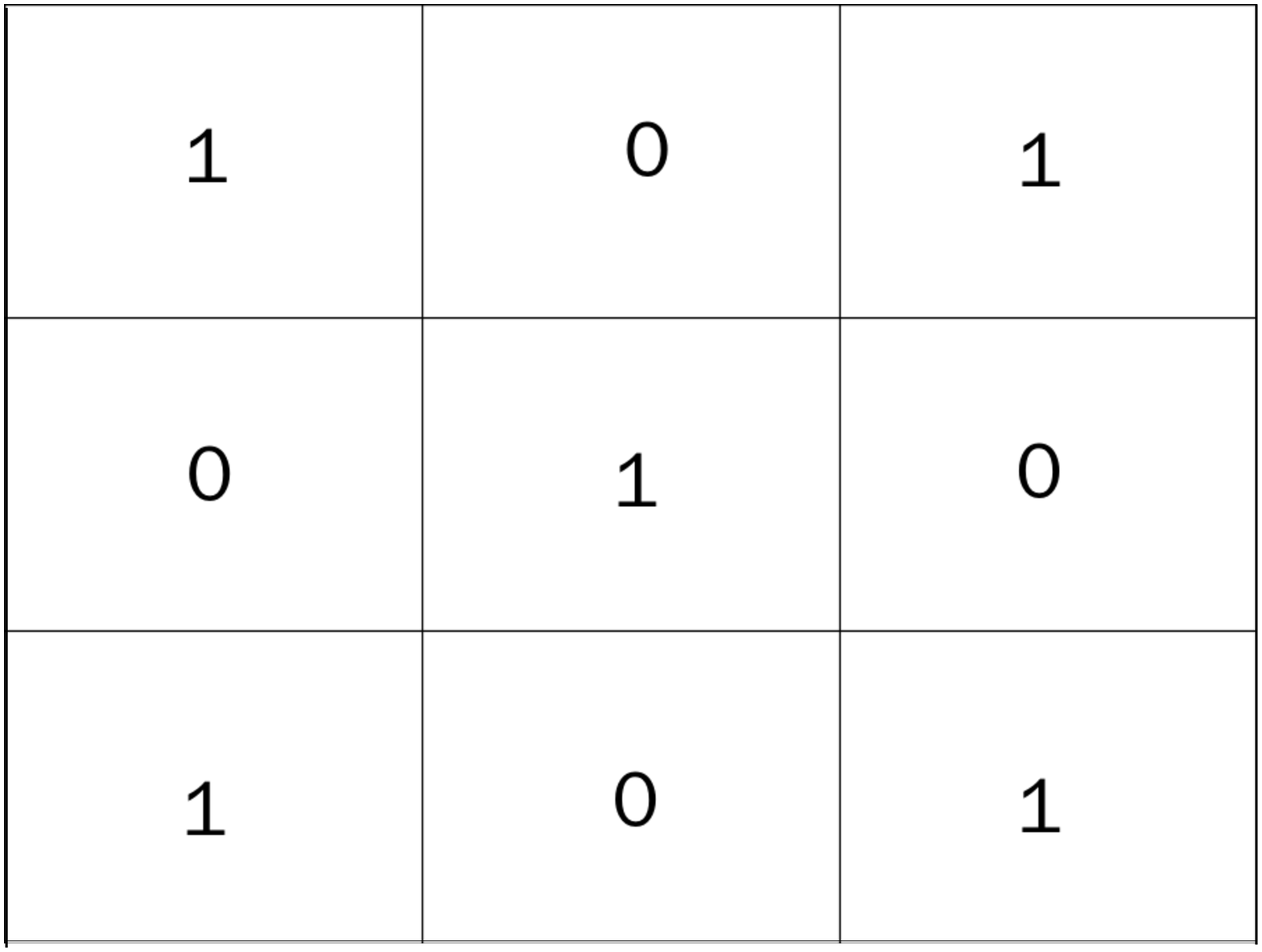}
\caption{This picture illustrates the spatial distribution of the part of SEP source
 near the Sun.
The area of source is divided evenly in latitudes and longitudes. The regions filled with ions are marked as ``1''; the regions devoid of ions are marked as ``0''. \label{source}}
\end{figure}

The particle source  on the solar wind source surface  covers  certain
ranges of  longitudinal and latitudinal 
$S_{long}\times S_{lat}$.
The spatial distribution of the SEP source is shown in 
Figure \ref{source}. The  source region is divided into small cells 
with the same longitude and latitude intervals. The regions   filled with ions are 
labeled as ``1'', and the regions devoid of ions are labeled as ``0''.  The size of every cell is set as 
$1.5^\circ$ in both latitudinal and longitudinal direction.
This setup of source region is to mimic the effect of braided magnetic field lines due to random walk of foot-point in the low corona. The size of the cell is equivalent to the typical size of supergranular motion. 
Without perpendicular diffusion, the particles propagate outward from the source
 to the interplanetary space only  along  the field lines. In 
this case, only the magnetic flux tubes connected to the regions labelled as ``1'' in the phase of source particle injection are 
filled with particles, and the rest of tubes are devoid of  particles. 
As the magnetic flux tubes past by an observer at 1 AU, the observer can see alternating switch-ons and switch-offs of SEPs.
However, with 
perpendicular diffusion, particles can
 cross the field lines when they  propagate  in the interplanetary  space. In this 
case, the longitudinal gradients in the particle intensities at different locations 
of interplanetary space  will be reduced.

We use boundary values to model the particles' injection from the source. The source rotates with the Sun, and the boundary condition is chosen as the following form
\begin{equation}
{f_b}(z \le 0.05AU,{E_k},\theta ,\varphi ,t) = \frac{a}{t} \cdot \frac{{E_k^{ - \gamma }}}{{{p^2}}} \cdot \exp \left( { - \frac{{{t_c}}}{t} - \frac{t}{{{t_l}}}} \right) \cdot \xi,
\end{equation}
\begin{equation}
\xi\left( {\theta ,\varphi } \right)  = \left\{ {\begin{array}{*{20}{l}}
{{e^{( - a\phi /\phi_0  )}} {\kern 15pt}\textrm{in source region 1}}\\
{0 {\kern 15pt}\textrm{otherwise}}
\end{array}} \right.\nonumber \label{SEPsource}
\end{equation}
where the particles are injected from the SEP source near the Sun. $\xi $ indicates the spatial scale of every cell. $E_k$ is  the particles energy.  We set a typical value of
$\gamma=-3$ for the spectral index of source particles. Because of  adiabatic energy loss,  those particles observed at $1$ AU have less energy than their initial energy at the source.  In our simulations, energy of particles at source are just a few times larger than that of particles at $1$ AU.
Time constants  
$t_c=0.48$ hour and $t_l=1.24$  hours indicate the rise and decay
time scales, respectively.  The injection time scales are 
used to model an impulsive SEP event. $\phi$ is the
angle distance from the  center of the cell and where the particles are injected. $\phi_0$ and $a$ are the constant.  $\phi_0$ is set to be $0.75^\circ$ (the half width of each cell), but $a$ is allowed to change according to several different scenarios.
The inner boundary is 0.05 AU and the outer boundary is 50 AU.
The  transport equation (\ref{dfdt}) is
solved by a time-backward Markov stochastic process method
\citep{Zhang1999ApJ...513..409Z, Qin2006JGRA..11108101Q}.
The transport equation can be reformulated to stochastic differential equations, so it can be solved by a Monte-Carlo simulation of Markov
stochastic process, and the SEP distribution function can be derived.
In this method, we trace virtual particles from the observation point back to the injection time from the SEP source.
More details of the technique can be found in those references.

\section{RESULTS}

\subsection{ ${\kappa _ \bot }/{\kappa _\parallel }$ Ratio}

Figure \ref{compose_15_and_20_dBToB_03} shows the interplanetary magnetic field in the ecliptic in the left panel, and the omni-directional fluxes for $500$ keV protons which are detected at $1$ AU in the right panel.  
In the left panel, the grey region indicates that the field lines are  
connected to the source. The dropouts and cutoffs are interpreted 
as  the magnetic flux tubes which are alternately filled with and 
devoid of ions pass the spacecraft. 
In our model, the source rotates with the Sun. As a result, the magnetic flux tube which connects with the source also rotates with the source.
The  observer is located at 
$1$ AU ($x=0, y=1$) in the equatorial plane as indicated by the black circle. The magnetic flux tubes rotate with the Sun, and the angular speed is $0.55^\circ$ per hour. According to the typical size of supergranulation, the size of every cell is set as $1.5^\circ$ in both latitudinal and longitudinal direction. 
An observer in the ecliptic traverses a cell in nearly $2.7$ hours.  
Due to the connection of the magnetic flux tubes, the observer's field lines can connect to different regions which are alternately filled with and devoid of ions.   In the right panel, the source parameter $a$ is set as $0$, so the source intensity is uniform in every cell.  
The source width is $S_{long}=S_{lat}=18^\circ$.
When the particles are injected, the observer at $1$ AU is magnetically connected with the first boundary of the source region. This same magnetic connection is also verified for the other simulations, except the last one when the observer is located at larger distance.
In all the cases, the parallel mean free paths are the same 
(${\lambda _\parallel =0.087}$ AU), but the perpendicular diffusion coefficients, and subsequently the ratios of perpendicular diffusion coefficients to parallel ones, are set to several different values.
With perpendicular diffusion, particles can be detected even if the observer is not 
connected directly to  region 1 by field lines.
The observer detects enhancements of particles starting at nearly $0.3$ day after the particles are injected on the Sun. 
With a larger perpendicular diffusion, the 
onset time of flux changes to an earlier time. The onset time of flux is the 
earliest and the latest in the cases of ${\kappa_\perp}/{\kappa _\parallel } = 1 
\times  10^{-4} $ and ${\kappa _ \bot }/{\kappa _\parallel } = 0 $, respectively.  
During the time interval from $0.55$ to $0.65$ day, the observer's field lines are
 connected to  region 0.  Without perpendicular diffusion, the observer can not 
detect energetic particles, so the flux suddenly drops to zero.  As the 
${\kappa _ \bot }/{\kappa_\parallel }  $ increases from $0$ to  $1\times  10^{-4} $,
 the variation of flux  becomes increasingly smaller during the interval from $0.55$
 to $0.65$ day. Especially in the case of 
${\kappa_\bot}/{\kappa_\parallel}=1\times 10^{-4}$, there is essentially no difference 
in the flux  between the time intervals when the observer’s field lines are not 
connected to the source. Later on, when the observer is 
connected to the  region $0$  again during the time interval from $0.8$ to $0.9$ day, the  fluxes  behave similarly to those in the interval from  
$0.55$ to $0.65$ day. After $1.25$ day, the observer is completely disconnected from the source region. The flux decreases very quickly in the case of 
${\kappa_\bot}/{\kappa_\parallel } = 1 \times  10^{-5}$, but the decreases slow down 
as the perpendicular diffusion coefficient increases. 
The time between neighboring valleys and peaks is less than $3$
hours. Let us define a ratio $R_i=f_{p,i}/f_{v,i}$, where $f_{p,i}$ and 
$f_{v,i}$ are the ith peak value and valley value of the flux, respectively. If the 
ratio 
$R_i$ is larger than $2$, we count it as a dropout. 
Since the ratios $R_i$ in each of the valleys are similar,
we  only use the ratio $R_2$ to identify the
dropouts of the fluxes.
When ${\kappa _ \bot }/{\kappa _\parallel }$ is
set as $0$, $1 \times  10^{-5}$, and $5 \times  10^{-5}$,  $R_2$ is approximately 
equal to $ + \infty$, $30.4$, and $2.2$, respectively.
We find that  the dropouts can be  reproduced only in the cases with
${\kappa _ \bot }/{\kappa _\parallel } \lesssim 5 \times  10^{-5}$.
In any case, if a dropout is reproduced, a cutoff, the step-like 
intensities decrease without recovery, is also reproduced.  In this sense, dropouts
and cutoffs are the same phenomenon, with the only difference being whether or not
the flux is recovered
by a follow-on connection to the particle source.

\subsection{ Parallel Mean Free Path}
The level of parallel mean free path affects the speed of particle propagation from the Sun to the Earth.
Figure \ref{compose_15_and_20_dBToB_1} is 
similar to Figure \ref{compose_15_and_20_dBToB_03} but
with different turbulence parameters $A_1$ or parallel mean free paths.
The left panel  corresponds to  ${\lambda _\parallel } = 0.0{\rm{26}}$ AU, and 
the right panel is corresponding to ${\lambda _\parallel } = 0.{\rm{5}}$ AU.
Due to the smaller parallel mean free path in the left panel of  Figure 
\ref{compose_15_and_20_dBToB_1}, the onset  of the fluxes shifts to a later time than that in
 Figure \ref{compose_15_and_20_dBToB_03}. 
In the right panel, where the mean free path is larger than that in  
 Figure \ref{compose_15_and_20_dBToB_03}, the onset is earlier. 
In the left panel, the observer
only detects two dropouts, then its foot-point goes away 
becomes disconnected from  the source region.
However, due to a larger parallel mean free path, the observer detects four dropouts in the right panel. 
When ${\kappa _ \bot }/{\kappa _\parallel }$ is set as
 $1 \times  10^{-5}$  and $5 \times  10^{-5}$,  $R_2$ is approximately equal to  $560$  and $3.7$ in the left panel, and is approximately equal to  $5.5$ and  $1.5$ in the right panel, respectively. The dropouts
requires ${\kappa _ \bot }/{\kappa _\parallel } \lesssim 5\times  10^{-5}$ in the case of ${\lambda _\parallel } = 0.0{\rm{26}}$ AU, and requires ${\kappa _ \bot }/{\kappa _\parallel } \lesssim 1\times  10^{-5}$ in the case of ${\lambda _\parallel } = 0.{\rm{5}}$ AU. Otherwise, the dropouts disappear. The results show that the  upper limit  of  ${\kappa _ \bot }/{\kappa _\parallel }$  in dropouts changes little with different ${\lambda _\parallel}$. 
The relation of the
appearance of dropout and parallel mean free path at given  ${\kappa _ \bot }/{\kappa
_\parallel }$ ratio can be understood as follows. When the mean free path increase, it takes a shorter
time for the particles to propagate from the Sun to the Earth, during which the particle
can not diffuse across magnetic field lines too much even with a large perpendicular diffusion coefficient. Therefore,
it is the ratio of ${\kappa _ \bot }/{\kappa _\parallel }$ that determines whether a dropout of
particle intensity is observed at the distance of 1 AU.

\subsection{Spatial Variation of SEP Source}

\begin{figure}
\epsscale{1}
\plotone{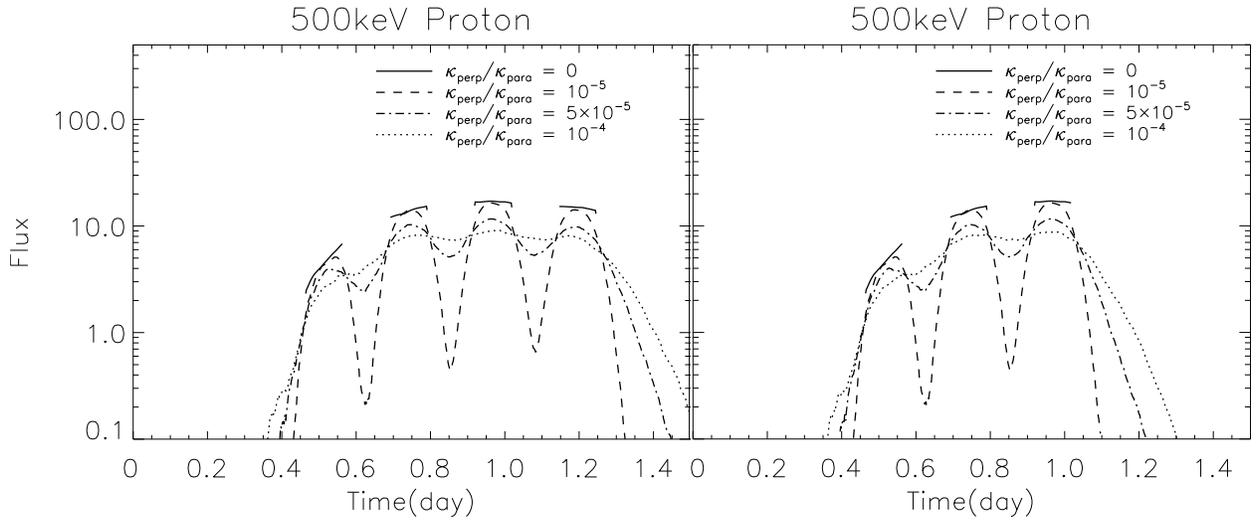}
\caption{
Comparison of $500$ keV proton fluxes with different source width.  
The width of source is $S_{long}=S_{lat}=18^\circ$ in the left panel,
and is $S_{long}=S_{lat}=15^\circ$ in the right panel. The $\lambda _\parallel$  is equal to $0.087$ $AU$, and  the source parameter $a$ is set as $0$ in all cases.
\label{different_source_width_dBToB_03_087AU}}
\end{figure}

The left panel of Figure \ref{different_source_width_dBToB_03_087AU} is the same as Figure \ref{compose_15_and_20_dBToB_03}. 
In the right panel, the source region is set to $S_{long}=S_{lat}=15^\circ$ which is narrower than that in the left panel.
Due to a narrower source, the observer  only encounters  two dropouts  instead of three.
Other than this, the fluxes in the right panel of Figure \ref{different_source_width_dBToB_03_087AU}  show behaviors similar to those in left panel.

\begin{figure}
\epsscale{1}
\plotone{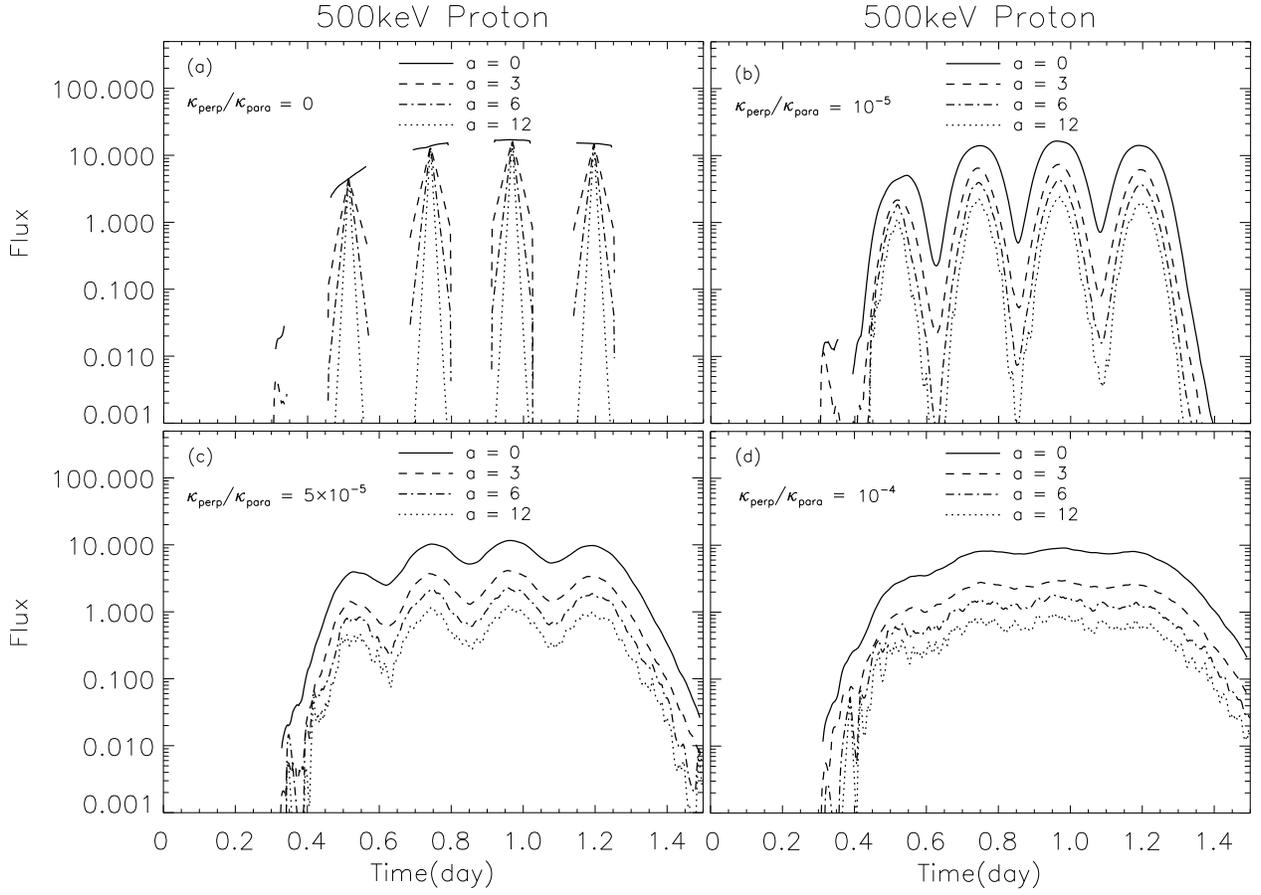}
\caption{
Comparison of $500$ keV proton fluxes with different spatial distribution of source. The source parameter $a$ changes from $0$ to $12$ in every panel. The four panels (a), (b), (c), and (d) are corresponding to $\kappa_\perp/\kappa _\parallel=0$, $\kappa_\perp/\kappa _\parallel=1 \times 10^{-5}$, $\kappa_\perp/\kappa _\parallel=5 \times 10^{-5}$, and $\kappa_\perp/\kappa _\parallel=1 \times 10^{-4}$, respectively. 
The $\lambda _\parallel=0.087$ AU  in the four panels. 
\label{different_source_spatial_variation_dBToB_03_087AU}}
\end{figure}

Figure \ref{different_source_spatial_variation_dBToB_03_087AU} shows  $500$ keV proton fluxes with different spatial distribution of source.  The four panels (a), (b), (c), and (d) correspond to $\kappa_\perp/\kappa _\parallel=0$, $\kappa_\perp/\kappa _\parallel=1 \times 10^{-5}$, $\kappa_\perp/\kappa _\parallel=5 \times 10^{-5}$, and $\kappa_\perp/\kappa _\parallel=1 \times 10^{-4}$, respectively. 
In every panel, the source parameter $a$ varies from $0$ to $12$. With a larger parameter $a$, the
source intensity decreases  more quickly towards the flank of each cell. In the panel (a), the ${\kappa _ \bot }/{\kappa _\parallel }$ is set as $0$. No particle is detected by the observer when the field line is disconnected from the source.
In the panel (b), the ${\kappa _ \bot }/{\kappa _\parallel }$ is set as $1 \times 10^{-5}$.
 Due to the variation of source intensity, the fluxes observed at 1 AU drop much more as $a$ increases. 
When $a$ increases from $0$ to $12$, $R_2$  also increase from 
$30.4$ to $1962$.
In the panel (c), the ${\kappa _ \bot }/{\kappa _\parallel }$ is set as $5 \times 10^{-5}$. The fluxes observed at 1 AU drop much slower as $a$ increases than that in the panel (b). When $a$ is set as $0$, $3$, $6$, and $12$, $R_2$ is approximately equal to $2.2$, $2.7$, $3.3$, and $3.6$.
In the panel (d),  the ${\kappa _ \bot }/{\kappa _\parallel }$ is set as $1 \times 10^{-4}$. There is no significant difference in the fluxes observed at 1 AU as $a$ 
increases. Based on the results in the four panels, we find that the dropouts can be  detected in the cases with  a slightly higher ratio of
${\kappa _ \bot }/{\kappa _\parallel } \lesssim 5\times  10^{-5}$, if the source distribution becomes narrower.

\subsection{Energy Dependence of Dropouts and Cutoffs}

\begin{figure}
\epsscale{1}
\plotone{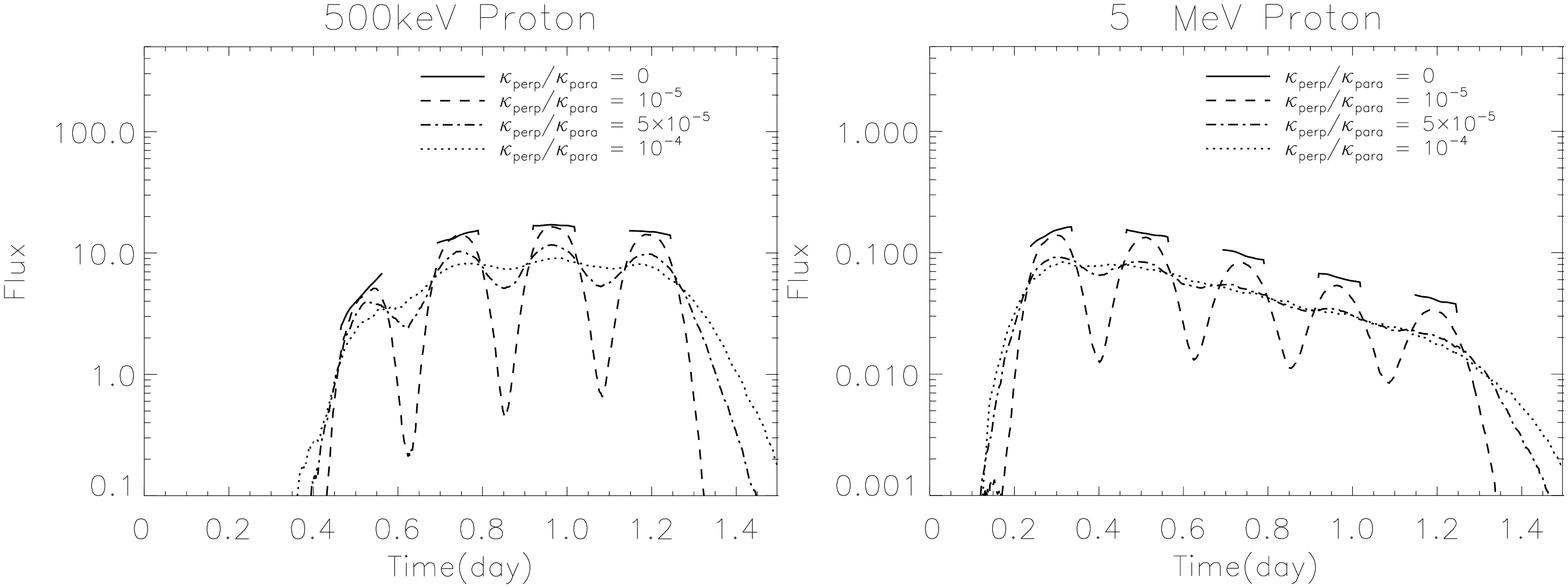}
\caption{
Comparison of  500 keV proton fluxes (the left panel) with 5 MeV proton fluxes (the right panel). 
 The source parameter $a$ is set as $0$ in the two panels.
\label{different_energy_channel_source_constant}}
\end{figure}

The left panel of Figure \ref{different_energy_channel_source_constant} is the same as Figure \ref{compose_15_and_20_dBToB_03}.
The right panel of Figure \ref{different_energy_channel_source_constant}  shows  the omni-directional flux for 5 MeV protons
detected at 1 AU in the ecliptic. The parallel  mean free paths remain the same 
(${\lambda _\parallel } = 0.{\rm{13}}$ AU at $1$ AU) in all cases, but the perpendicular diffusion coefficient is set to several  different  values. 
The source width in the two panels are set as  $S_{long}=S_{lat}=18^\circ$.
The source parameter $a$ is set as $0$ in the two panels.
Comparing with  the left panel, the onset  time of flux is earlier  because of the higher energy and  the larger parallel mean free path in the right panel.  In the right panel,  when ${\kappa _ \bot }/{\kappa _\parallel }$ is set as $0$, $1 \times  10^{-5}$, and $5\times 10^{-5}$, $R_2$ is approximately equal to $+\infty$, $10$, and $1.7$, respectively. The dropouts can be reproduced in the cases of 
${\kappa _ \bot }/{\kappa _\parallel } \lesssim 1 \times {10^{ - 5}}$ in the right panel. 
As a comparison, in the left panel, dropouts can be reproduced in the cases of
${\kappa _ \bot }/{\kappa _\parallel } \lesssim 5 \times {10^{ - 5}}$.

\subsection{Observer at A Larger Radial Distance}

\begin{figure}
\epsscale{1.0}
\plotone{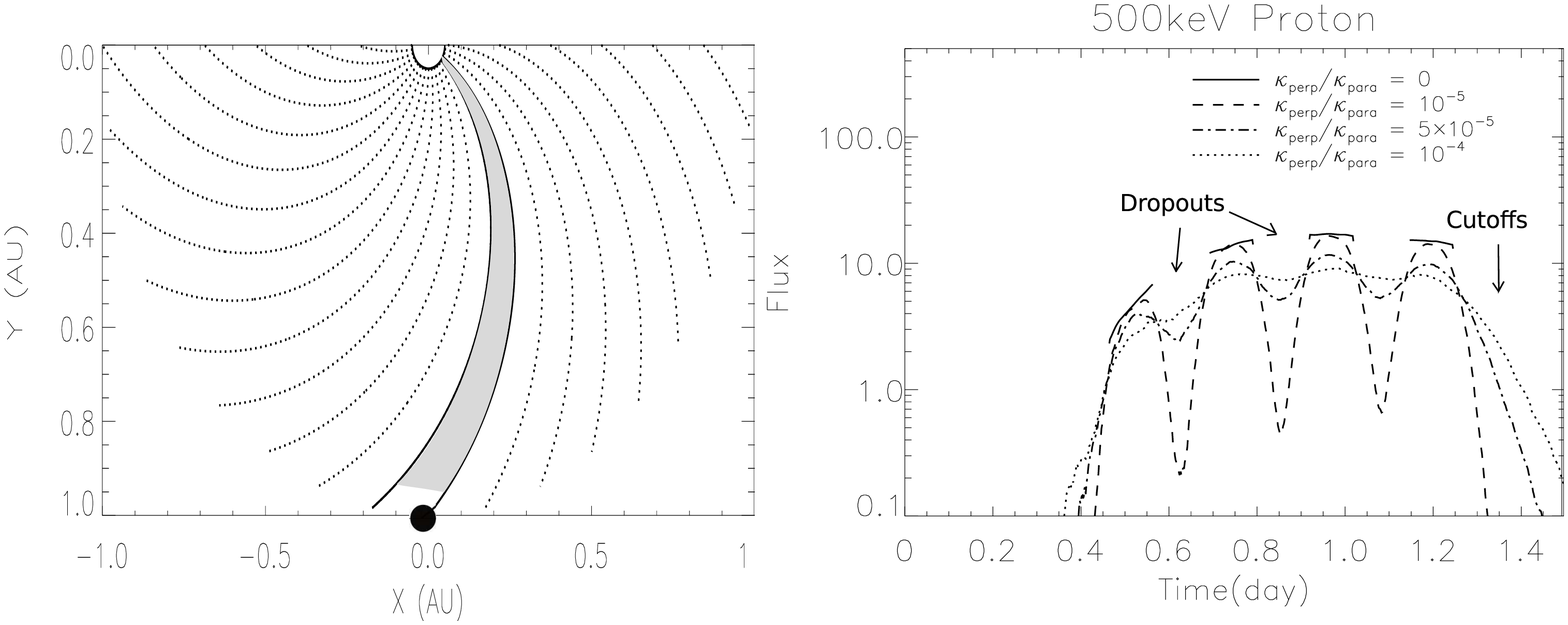}
\caption{
The left panel illustrates the spatial distribution of magnetic field lines in the interplanetary space. The area of source is divided evenly in latitudes and longitudes. The interplanetary magnetic field is set as the Parker model.The grey region indicates that the field lines connected to the source.  The observer is located at $1$ AU in the ecliptic as indicated by the black circle. In the right panel,  $500$ keV proton fluxes with different perpendicular diffusion are calculated. The $\lambda _\parallel$  is equal to $0.087$ $AU$, and  the source parameter $a$ is set as $0$ in all cases. Different lines correspond to the different ratios of
perpendicular diffusion coefficient to the parallel one: 
$\kappa_\perp/\kappa _\parallel=0$ (solid line); 
$\kappa_\perp/\kappa _\parallel=1 \times 10^{-5}$ (dash line); 
$\kappa_\perp/\kappa _\parallel=5 \times 10^{-5}$ (dash dot line); and 
$\kappa_\perp/\kappa _\parallel=1 \times 10^{-4}$ (dot line).
\label{compose_15_and_20_dBToB_03}}
\end{figure}

\begin{figure}
\epsscale{1}
\plotone{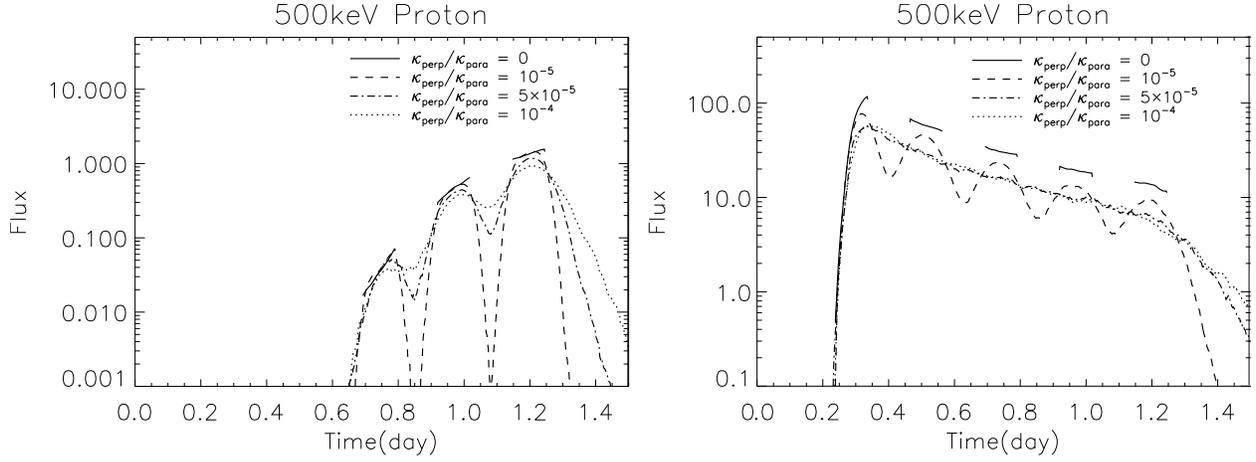}
\caption{
Similar to Figure \ref{compose_15_and_20_dBToB_03} but with different diffusion 
coefficients. The $\lambda _\parallel$ is $0.026$  AU in the left panel, and is $0.5$  AU in the right panel. The source parameter $a$ is set as $0$ in all cases.
\label{compose_15_and_20_dBToB_1}}
\end{figure}

\begin{figure}
\epsscale{1}
\plotone{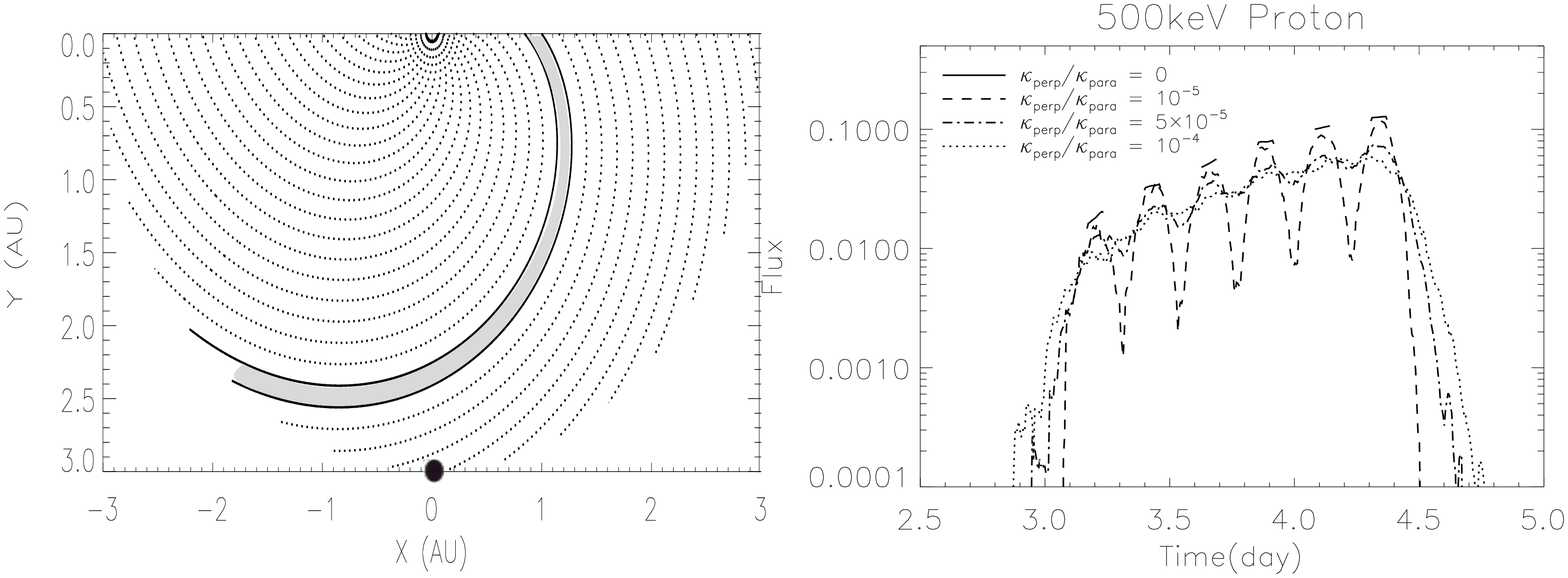}
\caption{
The left panel illustrates the spatial distribution of magnetic field lines in the interplanetary space. The observer is located
at 3 AU in the ecliptic as indicated by the black circle. In the right panel,  $500$ keV proton fluxes with different perpendicular diffusion are caculated.  The $\lambda _\parallel$  is equal to $0.087$ $AU$, and  the source parameter $a$ is set as $0$ in all cases.
\label{different_location_1AU_and_3AU_dBToB_03_087AU}}
\end{figure}

Results for an observer at $3$ AU in the ecliptic are shown in the 
Figure \ref{different_location_1AU_and_3AU_dBToB_03_087AU}.
The left panel  illustrates 
the interplanetary magnetic field lines. Due to a larger radial distance,
the particles spend more time propagating from the source
to the observer than  in the case of $1$ AU. In order to detect the particles,  the foot-point of observer is set as west $40^\circ$ to the boundary of source at the beginning of the simulation. In the right panel, the onset of the fluxes shifts to a later time than that in Figure \ref{compose_15_and_20_dBToB_03}. 
As the source rotates with the Sun, the observer  encounters  five dropouts, which is more than seen in Figure \ref{compose_15_and_20_dBToB_03}, because the observer is located at the boundary of the source  at the initial time in  Figure \ref{compose_15_and_20_dBToB_03}, and  particles spend some time propagating from source to $1$ AU. As a results, the observer missed two dropouts  in  Figure \ref{compose_15_and_20_dBToB_03}. In the right panel
of Figure \ref{different_location_1AU_and_3AU_dBToB_03_087AU}, the dropouts can be detected in the cases of  ${\kappa _ \bot }/{\kappa _\parallel } \lesssim 1 \times {10^{ - 5}}$.   In the case of  ${\kappa _ \bot }/{\kappa _\parallel } = 5 \times {10^{ - 5}}$, the dropout is absent in Figure \ref{different_location_1AU_and_3AU_dBToB_03_087AU}, but it
appears in Figure \ref{compose_15_and_20_dBToB_03}. 
The reason is that it takes a longer time for the particles to propagate from the Sun to the observer, and for perpendicular diffusion to be effective, when the solar radial distance increases.

\section{DISCUSSIONS AND CONCLUSIONS}
By numerically solving the Fokker-Planck focused transport equation for $500$ keV and $5$ MeV  protons,
we have investigated  the effect of the perpendicular diffusion coefficients on the dropouts and cutoffs  when an observer is located at $1$ AU or $3$ AU in the ecliptic.  SEPs are injected from a source near the Sun, and the source rotates with the Sun.
The dropouts and cutoffs are  caused by the magnetic flux tubes which are 
alternately filled with and devoid of ions past the spacecraft. In this paper, 
all the times between neighbouring valleys and peaks are less than $3$ hours, and the ratio $R_2$ between the second peak value and the second valley value are used to identify the 
dropouts. The dropout is defined to be present when  $R_2$ is
 more than a significant factor, which is set to be 2.
We list the values of  $R_2$ in the cases of different magnetic field turbulence intensities in Table \ref{diffusionCoefficients}.

Our simulations are performed for several different parallel mean free paths (${\lambda _\parallel } = $ $0.5$ AU, $0.087$ AU, $0.026$ AU at $1$ AU) with different  assumption for the ratios of perpendicular diffusion coefficient to the parallel one.  With a larger parallel mean free path, the onset 
time of SEP flux appears earlier, and more dropouts can be detected. 
Meanwhile, the flux increases more quickly, and the peak time is also earlier. This feature is closely related to the pitch angle distribution of  particles arriving at the observer. Since the particles encounter fewer scatterings when they propagate in the interplanetary space with a larger parallel mean free path. Therefore, the distribution of  pitch angle  would be anisotropic  for a longer time in this case.  
In order to reproduce the  dropouts and cutoffs at 1 AU, the perpendicular diffusion has to be small: ${\kappa _ \bot }/{\kappa _\parallel } \lesssim  5 \times {10^{ - 5}}$ when observer is located at $1$ AU, while ${\kappa _ \bot }/{\kappa _\parallel } \lesssim  1 \times {10^{ - 5}}$ when observer is located at $3$ AU. In any case when the dropouts are reproduced, cutoffs
can also be reproduced when the observer's flux tubes completely move out of the source region. If the observer is located at
a larger radial distance (eg, $3$ AU in our simulation), it takes a longer time for the particles to propagate from the Sun to the observer, and perpendicular diffusion has more time to be effective. In order to reproduce the dropout at several AU, the ratio of ${\kappa _ \bot }/{\kappa _\parallel }$ should be lower than that in the cases of $1$ AU. As a result, our simulation also predicts that the dropout may  disappear at larger radial distances, which can be checked by analysing data from Ulysses or other  spacecraft at large distance.

With a wider source, the observer can detect more
dropouts. Other than this, the fluxes with a wider source show behaviors similar to
 those with a narrower source. 
As $a$ changes from $0$ to $12$, in the case of ${\kappa_\bot}/{\kappa_\parallel}=1\times{10^{-5}}$, the ratio  $R_i$ is much larger in the case of $a=12$ than that in the case of $a=0$. However, in the cases of ${\kappa_\bot}/{\kappa_\parallel}=1\times{10^{-4}}$ and ${\kappa_\bot}/{\kappa_\parallel}=5\times{10^{-5}}$, the  ratio  $R_i$ does not change significantly as $a$ changes from $0$ to $12$.

For $5$ MeV protons, the case of ${\lambda _\parallel } = 0.13$ AU  has been 
analysed.  Due to  the higher particle speed and  a typically  larger parallel mean free path, the onset
 time appears earlier than for $500$ keV protons. As a result,
more dropouts can be detected. In order to reproduce the dropouts and cutoffs, the
ratio of the perpendicular diffusion coefficient to the parallel one should be 
smaller than $10^{-5}$, which is a little lower than that in the cases for $500$ 
keV protons.

Different ${\kappa _ \bot }$ and ${\kappa _\parallel } $ are 
obtained by altering the parameters $A_1$ and $A_2$ in our simulations, respectively, where 
 ${A_1}={\left({\delta{B_{slab}}}\right)^2}/({{{B_0}}^2}\cdot l_{slab})$, and 
${A_2}={\left({\delta{B_{2D}}}\right)^2}/{\left({{B_{0}}}\right)^2}\cdot{l_{2D}}$. 
In Table \ref{diffusionCoefficients}, we list all coefficients in the
diffusion formulae which are used in our simulations with $\lambda_\parallel$ equal
to $0.5$ AU, $0.087$ AU, $0.026$ AU for $500$ keV protons and $0.13$ AU for $5$ MeV protons at $1$ AU.
In this table, we assume a slab 
turbulence correlation length $l_{slab}=0.03$ AU, and 2D correlation length 
$l_{2D}=0.003$ AU. As we can see, the ${\left({\delta{B_{slab}}/{B_0}}\right)^2}$ is
 much larger than ${\left({\delta{B_{2D}}/{B_0}}\right)^2}$ in all cases. This result is consistent with the observation \citep{Tan2014Correlation}.  
However, we should note that the  exact values of 
${\left({\delta {B_{slab}}/{B_0}}\right)^2}$, 
${\left({\delta{B_{2D}}/{B_0}}\right)^2}$, $l_{slab}$ and $l_{2D}$ cannot be well 
determined. For example, we can assume slab turbulence correlation length 
$l_{slab}=0.003$ AU instead, and  ${\lambda _\parallel } = 0.5$ AU, $0.087$ AU, and 
$0.026$ AU for $500$ keV protons given 
${\left({\delta{B_{slab}}/{B_0}}\right)^2}=0.05$, 
$0.3$, and $1$, respectively. In our results we need  a very small
$\delta B_{2D}/\delta B_{slab}$ to get a small perpendicular diffusion coefficients  from the NLGC theory. 
However, the NLGC results with a small $\delta B_{2D}/\delta B_{slab}$ are much 
larger than simulation results \citep[e.g.,][]{Qin07}. Although 
$\delta B_{2D}/\delta B_{slab}$ must be small to get the small perpendicular 
diffusion coefficients, the actual value of $\delta B_{2D}/\delta B_{slab}$ needed 
according to simulations \citep{Qin07} is not as extremely small as that shown in 
Table \ref{diffusionCoefficients} from NLGC theory.

In \cite{Droge2010}, the cutoffs can be reproduced for a ratio of ${\kappa _ \bot }/{\kappa _\parallel }$ a few times $10^{-5}$.
This ratio is similar to what we deduced from our simulations. 
The basic difference between this simulation and the one
in \cite{Droge2010}  is that we reproduced the dropouts and cutoffs simultaneously, while their simulation only 
reproduced the cutoffs.
We believe that the cutoffs are only a special type of  dropout 
in which the intensity suddenly decreases without recovery.
In \cite{giacalone2000small} and \cite{Guo2013Small},  perpendicular diffusion coefficients relative to the background magnetic field (instead of Parker spiral) needed to be very small for reproducing the dropouts. That is consistent with our results. It should be noted that in our model, the large scale magnetic field is assumed to be a Parker spiral so that the Fokker-Planck focused  transport  equation can be  solved  efficiently with our stochastic method. In reality, the magnetic field lines with randomly walking foot-point do not have an azimuthal symmetry. However the large-scale geometry of interplanetary magnetic field and the behaviours of particle transport in it are not much different from those with a Parker magnetic field. The only difference is a slight shift of SEP source location relative to the magnetic field line passing through the observation at 1 AU.

Some special sets of turbulence parameters are needed in our simulations to 
produce  small perpendicular diffusion coefficients in order to produce the
 dropouts and cutoffs. For example, turbulence dominated by  a  
slab component leads to very small perpendicular diffusion coefficients. 
In the future, it will be interesting for us to study solar wind turbulence 
geometry from spacecraft observations when SEP dropouts and cutoffs occur.

\acknowledgments 
The authors thank the anonymous referee for valuable comments. Y. W.  benefited from the discussions with Fan Guo. We are partly supported by grants NNSFC 41304135,  NNSFC 41374177, and  NNSFC 
41125016, the CMA grant GYHY201106011, and the Specialized Research Fund for State 
Key Laboratories of China.
The computations were performed by Numerical Forecast Modeling R\&D
and VR System of State Key Laboratory of Space Weather and Special
HPC work station of Chinese Meridian Project.  M.Z. was supported in part by NSF under
Grant AGS-1156056 and by NASA under Grant NNX08AP91G. 
SD acknowledges funding from the UK Science and Technology Facilities 
Council (STFC) (grant ST/J001341/1) and the International Space Science 
Institute.


\begin{table}
\begin{centering}
\caption {Diffusion coefficients and flux ratios.\label{diffusionCoefficients}}
\begin{tabular} {|l|l|l|l|l|l|l|} \tableline

${\lambda _\parallel }$ $(AU)$&  $l_{slab}$ $(AU)$ & $l_{2D}$ $(AU)$  & ${\left( {\delta {B}/{B_0}} \right)^2}$ & ${\kappa_ \bot }/{\kappa _\parallel }$ &  ${\left( {\delta {B_{2D}}/{B_{slab}}} \right)^2}$ & ${R_2}^c$
\\ \tableline\tableline

\multirow{4}{*}{$0.5^a$ } & \multirow{4}{*}{$3 \times {10^{ - 2}}$ }   & \multirow{4}{*}{$3 \times {10^{ - 3}}$ }  & \multirow{4}{*}{0.05}    &  0 &  0  & $  + \infty $ \\
&  &  &  &  $1 \times 10^{-5} $  &  $1 \times 10^{-5}$  & $ 5.5$ \\
&  &  &  &  $5 \times 10^{-5} $  &  $9 \times 10^{-5}$  & $ 1.5$ \\
&  &  &  &  $1 \times 10^{-4} $  &  $3 \times 10^{-4}$  & $ 1.5$ \\
\tableline\tableline

\multirow{4}{*}{$0.087^a$ } & \multirow{4}{*}{$3 \times {10^{ - 2}}$ }   & \multirow{4}{*}{$3 \times {10^{ - 3}}$ }  & \multirow{4}{*}{0.3}    &  0 &  0 & $  + \infty $  \\
&  &  &  &  $1 \times 10^{-5} $  &  $2 \times 10^{-6}$  & $ 30.4$ \\
&  &  &  &  $5 \times 10^{-5} $  &  $2 \times 10^{-5}$  & $ 2.2 $ \\
&  &  &  &  $1 \times 10^{-4} $  &  $5 \times 10^{-5}$  & $ 1.1$ \\
\tableline\tableline

\multirow{4}{*}{$0.026^a$ } & \multirow{4}{*}{$3 \times {10^{ - 2}}$ }   & \multirow{4}{*}{$3 \times {10^{ - 3}}$ }  & \multirow{4}{*}{1}    &  0 &  0 & $  + \infty $ \\
&  &  &  &  $1 \times 10^{-5} $  &  $4 \times 10^{-7}$  & $ 560$\\
&  &  &  &  $5 \times 10^{-5} $  &  $5 \times 10^{-6}$  & $ 3.7$\\
&  &  &  &  $1 \times 10^{-4} $  &  $1 \times 10^{-5}$  & $ 1.4$\\
\tableline\tableline

\multirow{4}{*}{$0.13^b$ } & \multirow{4}{*}{$3 \times {10^{ - 2}}$ }   & \multirow{4}{*}{$3 \times {10^{ - 3}}$ }  & \multirow{4}{*}{0.3}    &  0 &  0 & $  + \infty $ \\
&  &  &  &  $1 \times 10^{-5} $  &  $2 \times 10^{-6}$  & $ 10$\\
&  &  &  &  $5 \times 10^{-5} $  &  $2 \times 10^{-5}$  & $ 1.7$\\
&  &  &  &  $1 \times 10^{-4} $  &  $7 \times 10^{-5}$  & $ 1.3$\\
\tableline

\end{tabular}
\tablenotetext{a}{For $500$ keV protons at $1$ AU.}
\tablenotetext{b}{For $5$ MeV protons at $1$ AU.}
\tablenotetext{c}{The source parameter $a$ in equation \ref{SEPsource} is set as 0, and the observer is located at $1$ AU.}
\end{centering}
\end{table}

\end{document}